\documentclass[journal]{IEEEtran}

\usepackage{algpseudocode}
\usepackage{amsfonts}       
\usepackage{amsmath}
\usepackage{array}
\usepackage{booktabs}       
\usepackage{enumitem}
\usepackage[T1]{fontenc}    
\usepackage{graphicx}
\usepackage{hyperref}       
\usepackage[utf8]{inputenc} 
\usepackage{microtype}      
\usepackage{nicefrac}       
\usepackage{subcaption}
\usepackage{url}            
\usepackage[numbers,sort&compress]{natbib}
\usepackage[dvipsnames]{xcolor}

\usepackage{listings}
\usepackage{color}
\usepackage{pifont}
\newcommand{\cmark}{\ding{51}}%
\newcommand{\xmark}{\ding{55}}%

\newcolumntype{L}[1]{>{\raggedright\let\newline\\\arraybackslash\hspace{0pt}}m{#1}}
\newcolumntype{C}[1]{>{\centering\let\newline\\\arraybackslash\hspace{0pt}}m{#1}}
\newcolumntype{R}[1]{>{\raggedleft\let\newline\\\arraybackslash\hspace{0pt}}m{#1}}

\usepackage{tikz}

\begin{document}

\title{MRI Pulse Sequence Integration for Deep-Learning Based Brain Metastasis Segmentation}

\author{Darvin~Yi*,
        Endre~Gr{\o}vik,
        Michael~Iv,
        Elizabeth~Tong,
        Kyrre~Eeg~Emblem,
        Line~Brennhaug~Nilsen,
        Cathrine~Saxhaug,
        Anna Latysheva,
        Kari~Dolven~Jacobsen,
        {\AA}slaug~Helland,
        Greg~Zaharchuk,
        and~Daniel~Rubin
\thanks{\textit{Asterisk indicates corresponding author.}}
\thanks{D. Yi* is with the Department of Biomedical Data Science at Stanford University, Stanford, CA 94305 USA e-mail: darvinyi[at]Stanford[dot]EDU}
\thanks{E. Gr{\o}vik, K. E. Emblem, and L. B. Nilsen are with the Department for Diagnostic Physics at Oslo University Hospital}
\thanks{M. Iv, E. Tong, and G. Zaharchuk are with the Department of Radiology at Stanford University.}
\thanks{C. Saxhaug and A. Latysheva are with the Department of Radiology and Nuclear Medicine at Oslo University Hospital.}
\thanks{K. D. Jacobsen and {\AA}. Helland is with the Department of Oncology at Oslo University Hospital.}
\thanks{D. Rubin is with the Department of Biomedical Data Science and Department of Radiology at Stanford University.}}

\markboth{DRAFT UPDATED 18 DECEMBER 2019}%
{Yi \MakeLowercase{\textit{et al.}}: Magnetic Resonance Pulse Sequence Integration for Metastasis Segmentation}

\maketitle

\begin{abstract}
Magnetic resonance (MR) imaging is an essential diagnostic tool in clinical medicine.  Recently, a variety of deep learning methods have been applied to segmentation tasks in medical images, with promising results for computer-aided diagnosis.  For MR images, effectively integrating different pulse sequences is important to optimize performance.  However, the best way to integrate different pulse sequences remains unclear.  In this study, we evaluate multiple architectural features and characterize their effects in the task of metastasis segmentation.  Specifically, we consider (1) different pulse sequence integration schemas, (2) different modes of weight sharing for parallel network branches, and (3) a new approach for enabling robustness to missing pulse sequences.  We find that levels of integration and modes of weight sharing that favor low variance work best in our regime of small data (n = 100).  By adding an input-level dropout layer, we could preserve the overall performance of these networks while allowing for inference on inputs with missing pulse sequence.  We illustrate not only the generalizability of the network but also the utility of this robustness when applying the trained model to data from a different center, which does not use the same pulse sequences.  Finally, we apply network visualization methods to better understand which input features are most important for network performance.  Together, these results provide a framework for building networks with enhanced robustness to missing data while maintaining comparable performance in medical imaging applications.
\end{abstract}


\IEEEpeerreviewmaketitle

\begin{figure}[htb]
	\centering
	\includegraphics[width=0.45\textwidth]{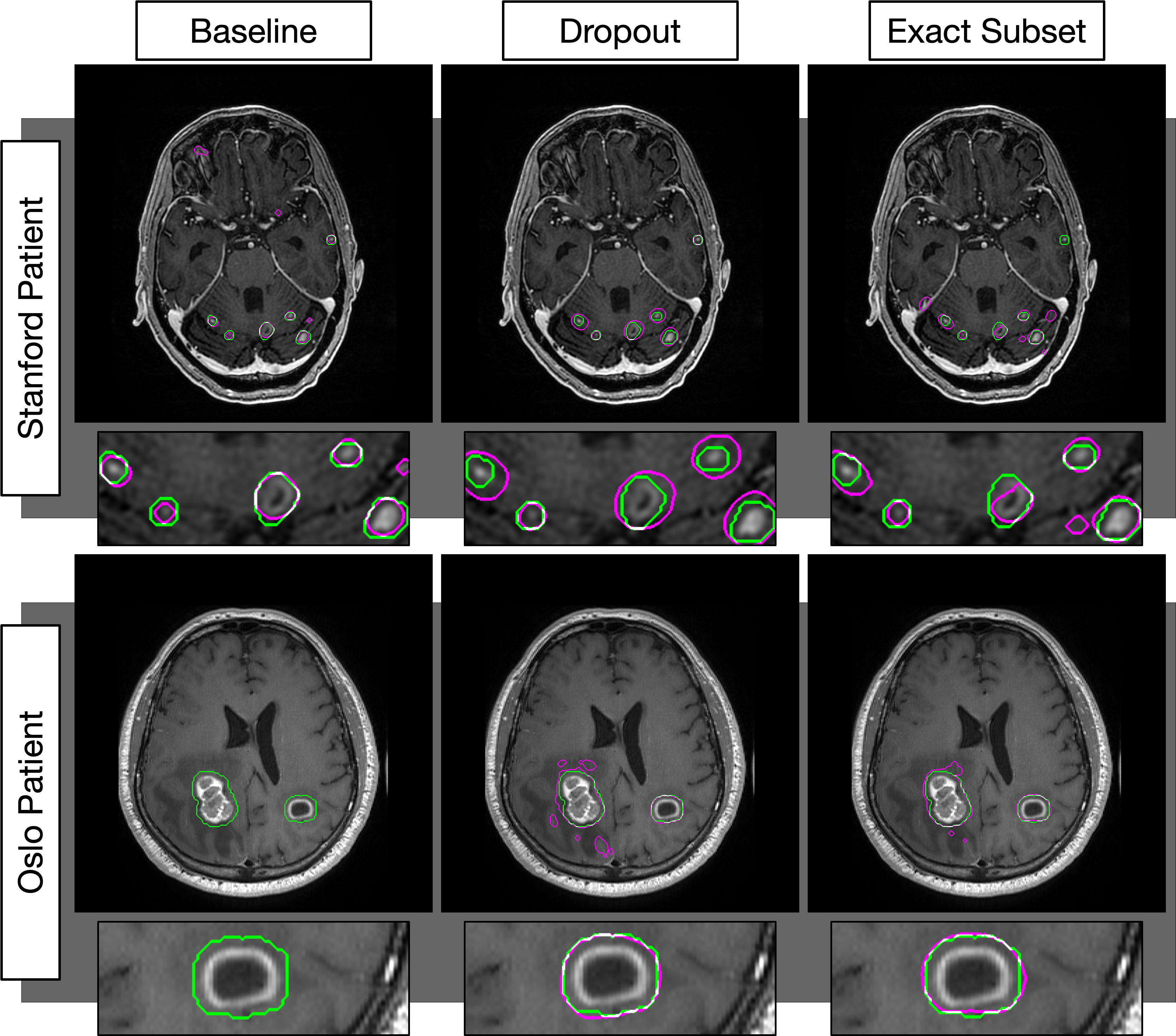}
	\caption{\textbf{Visualization of Segmentations on Stanford and Oslo Patients.}  \small{This figure shows visualizations of the results from our baseline input-level integration network, the dropout trained equivalent, and the baseline network trained (and tested) on the BRAVO-censored data.  Our networks were only trained on Stanford patients, and we show examples for a Stanford patient and an Oslo patient, both not used during training.  Below each image, we show a magnified zoom on areas of possible interest.  Expert Annotation segmentations are outlined in green.  Predictions are outlined in purple/magenta.  Overlap will be shown as white.}}
	\label{fig:visualization}
\end{figure}

\section{Introduction}

\IEEEPARstart{R}{}ecently, there has been an explosion in applying deep learning methods to magnetic resonance (MR) imaging.  Many techniques utilize classification networks to aid diagnosis and detection \cite{backstrom2018efficient,lee2017deep,zeng2018multi}. Other techniques applied the generative capabilities of convolutional neural networks (CNNs) - e.g. through autoencoder networks \cite{doersch2016tutorial,tolstikhin2017wasserstein,vincent2008extracting,wang2013learning} or generative adversarial networks (GANs) \cite{arjovsky2017wasserstein,goodfellow2014generative} - to denoise MR images \cite{gondara2016medical} and enhance signal-to-noise ratio \cite{gong2018deep,zhu2018image}.  Finally, deep learning for segmentation has been of core interest to the computational biomedical community since its advent.  Biomedical segmentation produced U-Net \cite{ronneberger2015u}, an architecture used in the broader computer science community to this day.  For MR imaging, both 2.5D networks and 3D networks have been commonly used to solve segmentation problems \cite{bakas2018identifying,kamnitsas2016deepmedic,milletari2016v}.  In this work, we will focus on further developing segmentation techniques for MR images.

MR imaging protocols vary from institution to institution and case by case.  Typical brain metastasis MRI protocol consists of T2-weighted Fluid attenuation inversion recovery (FLAIR) sequence, pre-contrast and post-contrast T1-weighted sequences. The critical sequence is the post-contrast 3D T1-weighted sequence, which is a high-resolution isotropic sequence acquired by either Inversion Recovery prepped Fast Spoiled Gradient-Echo (IR-FSPGR) or Fast Spin-Echo (FSE) techniques. 3D T1-weighted IR-FSPGR acquire isotropic T1-weighted images with excellent grey-white matter differentiation and are used broadly. 
It is often difficult to distinguish enhancing metastases from background vascular enhancement on post-contrast 3D T1-weighted IR-FSPGR \cite{bae2017efficacy}. On the contrary, post-contrast 3D T1-weighted FSE provides inherent background vascular suppression, yielding a higher contrast-to-noise ratio (CNR) than post-contrast FSPGR, making enhancing metastases more conspicuous \cite{majigsuren2015comparison}. 
No matter the protocol, networks trained with a set of input pulse sequences are notoriously sensitive to missing pulse sequences at run-time \cite{havaei2016deep,havaei2016hemis,li2014deep,van2015does}.

In this work, we characterize what we have identified as best practices for pulse sequence integration.  We evaluate how to integrate multiple pulse sequences for optimal segmentation and detection performance.  We also introduce how using a ``dropout'' layer on the integration layer for our networks confers robustness to missing pulse sequences with no significant loss of performance given all pulse sequences.  Finally, we show that this integration dropout method gives a more consistent training schedule, or behavior of gradients, by analyzing input image saliency maps throughout training.

\section{Related Work}

\textbf{Deep Learning and Segmentation.}  Deep learning-based methods have shown excellent performance for segmentation.  Following the first fully convolutional network (FCN) in 2014 \cite{long2015fully}, semantic segmentation with deep learning has continued to improve with SegNet \cite{badrinarayanan2017segnet} and U-Net \cite{ronneberger2015u} in 2015, PSPNet \cite{zhao2017pyramid} and DeepLab \cite{chen2017deeplab} in 2017, and recently DeepLabv3 \cite{chen2017rethinking}.  Constant advances in computing power and use of contextual information have provided further benefits.

\textbf{Segmentation for MR Imaging.}  The MR community has a strong background in using deep segmentation techniques.  Open challenges like the BraTS \cite{bakas2018identifying} or the Promise12 \cite{litjens2014evaluation} challenges have provided many datasets to train segmentation algorithms in the MR domain.  Though earlier approaches typically use single-frame 2D FCN approaches \cite{zhang2015deep}, current approaches leverage 3D information in either 2.5D or 3D FCNs \cite{fritscher2016deep,xu2017neonatal,kamnitsas2016deepmedic}.  Other methods have investigated using deep learning methods from the recurrent network domain \cite{zhao2018deep}, a type of network built to handle sequence-like data.  These methods represent the variety of deep learning-based segmentation techniques that have proven valuable in the quantitative MR community.

\textbf{Multi-modal Data Integration.}  Because of deep network’s generally malleable architecture, many groups have investigated the best modes to combine different data modalities.  Our work on integrating different pulse sequences draws from data integration schemes first found in video frame analysis \cite{karpathy2014large}, which create parallel shared weights before the layer of integration, after which there is only one shared network.  In investigating parallel network branches, we also looked into the work done with siamese networks \cite{bertinetto2016fully,hoffer2015deep,wang2013learning}.  Most notably, our study closely models that of Eitel et. al. on training robust RGB-D networks \cite{eitel2015multimodal}.  In the MR space, \cite{van2015does} has been done to create synthetic modalities to replace missing data should the case arise.  However, most similar to our own work would be \cite{havaei2016hemis}'s 2016 Hetero-Modal Image Segmentation (HeMIS) where each present modality ``votes'' within a learned latent representation space by means of arithmetic mean.  Nonetheless, effectively addressing the problem of missing data remains an important outstanding challenge for MR images.

\textbf{Dropout Regularization.}  Dropout, an early addition to the deep learning toolbox, helps decrease learned co-linearities during training \cite{ba2013adaptive,srivastava2014dropout} by randomly setting activations to zero during training based on a predetermined probability.  An equivalent channel-wise dropout is used for training CNNs \cite{smirnov2014comparison}.

\textbf{Network Performance Visualization.}  Many techniques, such as the class action map (CAM), have been developed to help visualize network decision making with respect to the original input images \cite{zhou2016learning}.  These visualization methods have been quite popular in medical imaging applications as the main form of network explanation \cite{rajpurkar2017chexnet,rajpurkar2017mura}.  Here, we focus on the idea behind the saliency map, a gradient-based method \cite{hong2015online,simonyan2013deep}.

\textbf{Previous Work and Novel Contributions.}  Our group’s previous work set the baseline performance for our metastasis segmentation task \cite{grovik2019deep}.  In comparison, we have updated the segmentation architecture and used detection metrics to better describe our network’s performance.  Methodologically, we build on previous works in both medical imaging and real-world computer vision to investigate the best methods of integrating multi-modal MR pulse sequence data for metastasis segmentation.  In this paper, we report the following novel contributions:
\begin{enumerate}
    \item An initial investigation into different pulse sequence integration architectures and training techniques for metastasis segmentation in the small data regime.
    \item Creation and usage of an input-layer dropout that makes the network robust to receiving missing pulse sequences and introduces more consistent network training behavior.
    \item Evaluation of the performance of the trained network on different pulse sequences to show which input modalities are most and least useful.
\end{enumerate}

\section{Data}

This retrospective, multi-center study was approved by our Institutional Review Board.  For our experiments, we use brain MR data from two different institutions: Stanford Hospital and Oslo University Hospital. These datasets will hereafter be referred to as the Stanford data and the Oslo data.  For both cohorts, all MR image-series were co-registered into one common anatomical space. This was performed using the nordicICE software package (Nordic Neuro Lab, Bergen, Norway). For the Stanford data, all image series were co-registered to a post-contrast 3D T1-weighted inversion recovery fast spoiled gradient echo, whereas for the Oslo data, post-contrast 3D T1-weighted spin echo images was used at the reference series. Furthermore, for the Oslo data, a defacing procedure was applied to anonymize all image-data using an in-house algorithm (MATLAB R2017a version 9.2.0, MathWorks Inc. Natick, MA, USA).

\subsection{Stanford Metastasis Cohort}

A total of 156 patients with brain metastases were examined at Stanford Hospital. Inclusion criteria for patient enrollment were presence for known metastatic lesion(s), no prior surgical or radiation treatment, and the availability of the required MR images. Imaging was performed on both 1.5T (SIGNA Explorer and TwinSpeed, GE Healthcare, Chicago, IL) and 3T (Discovery 750 and 750w and SIGNA Architect, GE Healthcare, Chicago, IL; Skyra, Siemens Healthineers, Erlangen, Germany) clinical scanners. The imaging protocol included pre- and post-contrast T1-weighted 3D fast spin echo, post-Gd T1-weighted 3D axial inversion recovery prepped fast spoiled gradient-echo, and 3D fluid-attenuated inversion recovery. For contrast-enhancement, a dose of 0.1 mmol/kg body weight of gadobenate dimeglumine (MultiHance, Bracco Diagnostics, Princeton, New Jersey) was intravenously administered.  Primary cancers from the Stanford dataset consists of lung (99), breast (33), skin/melanoma (7), genitourinary (7), gastrointestinal (5), and other miscellaneous primary sources (5).

\subsection{Oslo Metastasis Cohort}

For the Oslo data, a total of 65 patients with brain metastases were examined. To be eligible for inclusion, patients had to receive stereotactic radiosurgery for at least one brain metastasis measured to a minimum of 5 mm in one direction, be untreated or progressive after systemic or local therapy, have confirmed non-small-cell lung cancer or malignant melanoma, be $\geq$18 years of age; have an Eastern Cooperative Oncology Group performance status score of maximum 1, and have a life expectancy of more than 6 weeks. All imaging was performed on a clinical 3T Skyra scanner (Siemens Healthineers, Erlangen, Germany). The imaging protocol included pre- and post-contrast T1-weighted 3D fast spin echo and 3D T2-weighted FLAIR. For contrast-enhancement, a dose of 0.1 mmol/kg body weight of gadoterate (Dotarem, Guerbert, France) was intravenously administered. Note that the Oslo data only had three of the four MR sequences acquired in the Stanford data, and that two patients in the Oslo cohort were missing the pre-contrast T1-weighted 3D fast spin echo. Hence, the data represents a real-world use case of having to create a model that would be robust to missing pulse sequences, a scenario quite common considering that different institutions use differing imaging protocols. The primary cancers of the Oslo data consists of lung (45) and melanoma (20).

\section{Methods}

\subsection{Multiple Levels of Pulse Sequence Integration}

As described above, our MR images are 3-D voxel representations of different physical properties captured in different pulse sequences.  Because they have been co-registered with each other, the voxel at location $[i,j,k]$ for each pulse sequence captures the same physical location.  Thus, pixel-wise operations, such as averaging or channelwise concatenation, are location-preserving and do not have to rely on operations such as flattening.

\begin{figure}[htb]
	\centering
	\begin{subfigure}[h]{0.45\textwidth}
		\includegraphics[width=\textwidth]{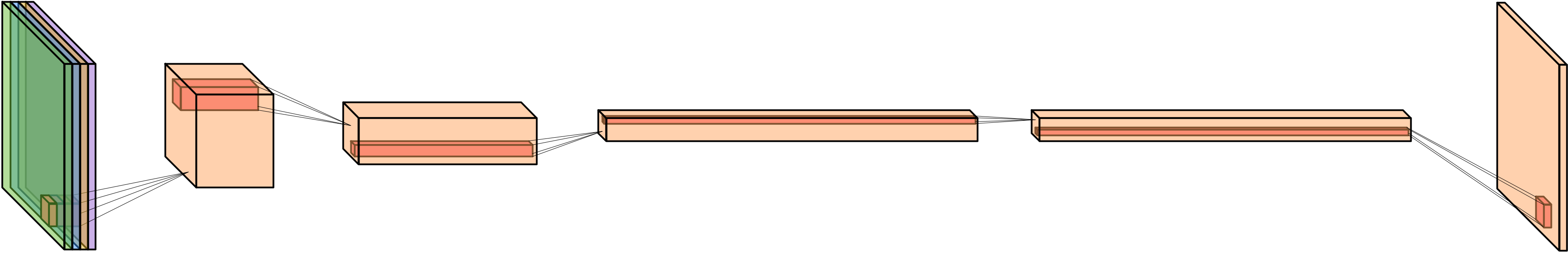}
		\caption{Input-Level Integration}
	\end{subfigure}
	~
	\begin{subfigure}[h]{0.45\textwidth}
		\includegraphics[width=\textwidth]{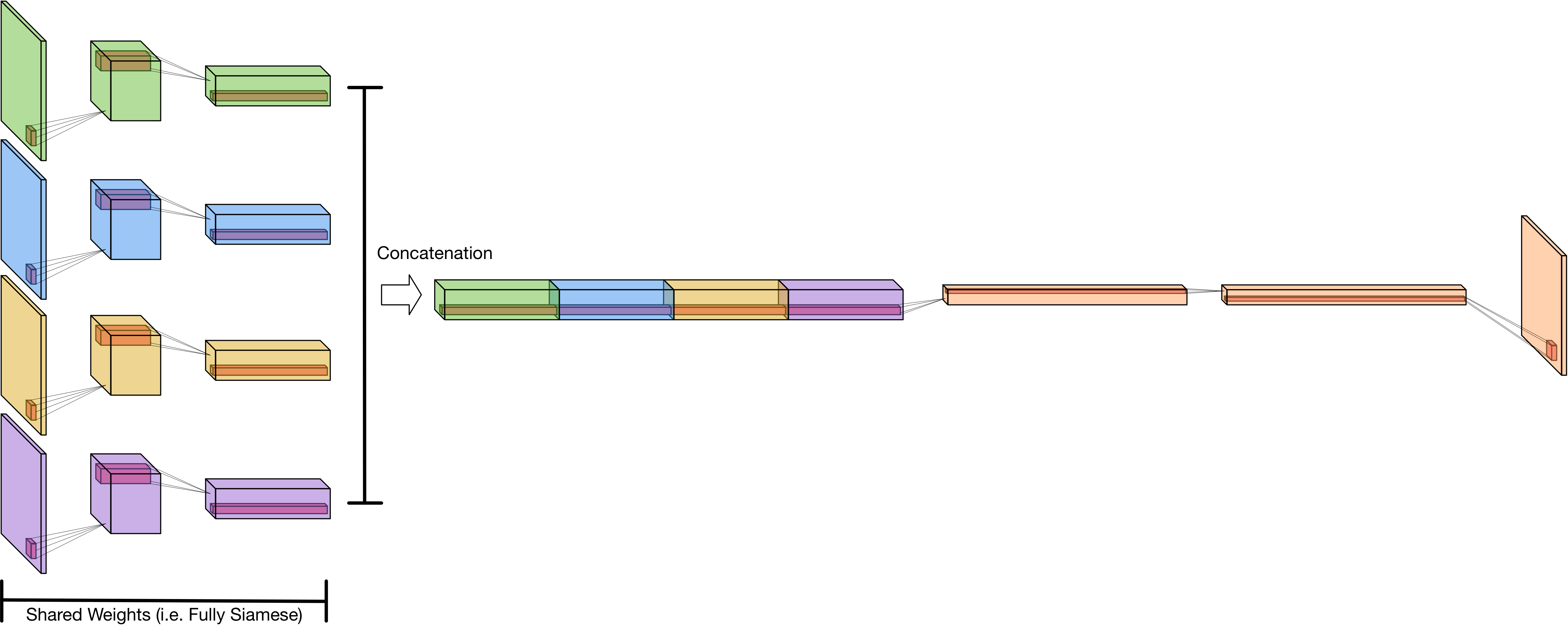}
		\caption{Mid-Level Integration}
	\end{subfigure}
	~
	\begin{subfigure}[h]{0.45\textwidth}
		\includegraphics[width=\textwidth]{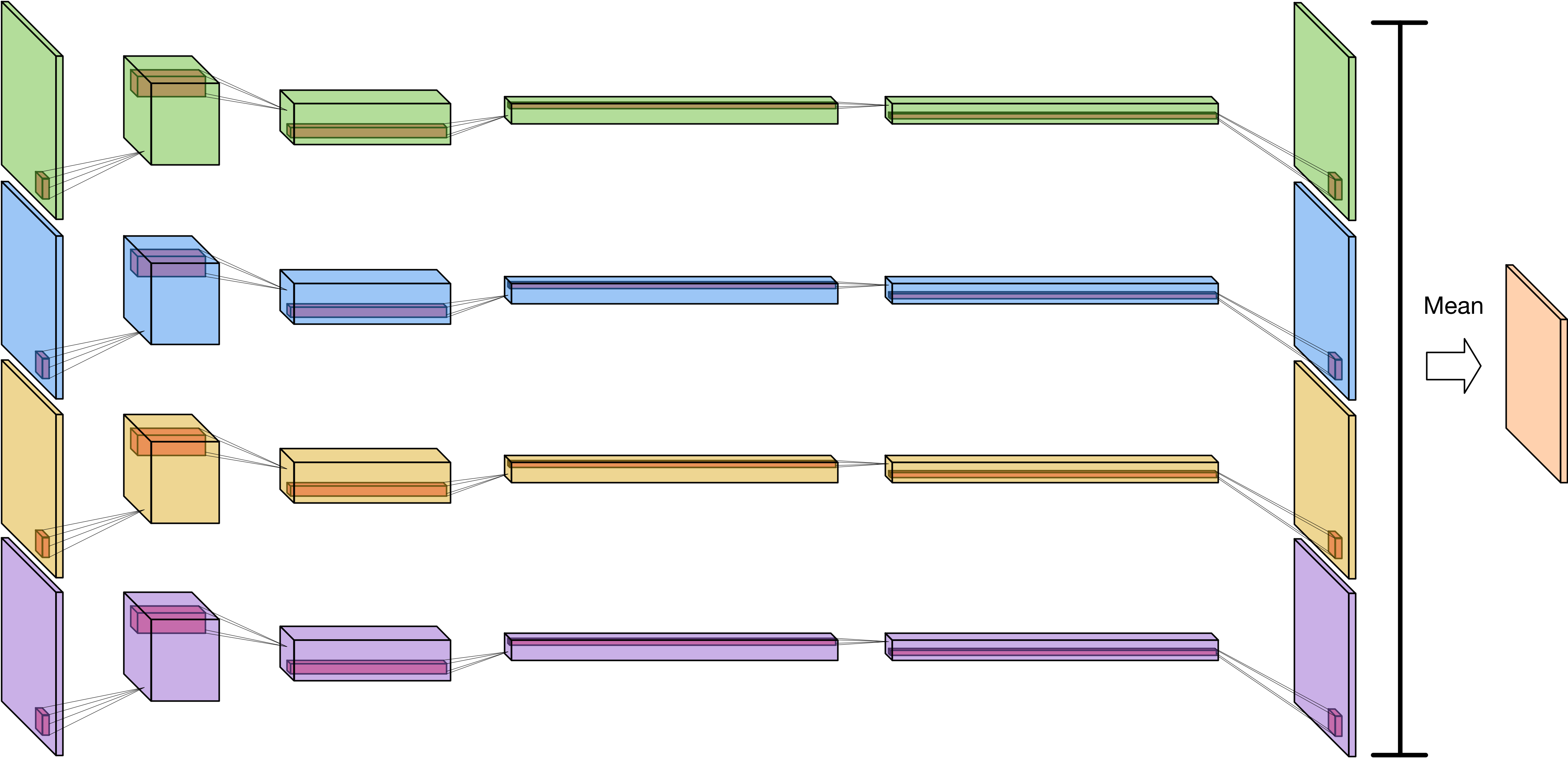}
		\caption{End-Level Integration}
	\end{subfigure}
	\caption{\textbf{Pulse Sequence Integration levels.} \small{(a) The baseline, as used in previous work \cite{grovik2019deep}, integrates the pulse sequence images at the input level.  (b) The mid-level integration has both separate and shared layers.  We do this integration after DeepLab v3’s block 4, before the atrous spatial pooling.  (c) End-level integration gives each pulse sequence a separate network and takes the mean of the final predictions.  All network diagrams are not representative of the actual convolutional architecture, but merely a placeholder to diagram the larger concept of multi-modal feature integration.}}
	\label{fig:integrationLevels}
\end{figure}

Common approaches to combine information from different branches of co-localized data are taking some weighted sum (e.g. average) or performing a concatenation in the channel space.  In our experiments, we use both of these methods.  An outstanding question for integrating information from each of the different pulse sequences is how much to process distinct pulse sequences in isolation and how much to process them after integration.  Different integration schemes have been explored in video frame analysis and other applications, but limited work has been done to characterize their performance for MR images. Although there exists an intractable number of integration schemes, we focus on three levels: (a) input-level integration, (b) mid-level integration, and (c) end-level integration.  These three schemes are diagrammed in Figure \ref{fig:integrationLevels}.

\subsection{Multiple Modes of Weight Sharing}

In addition to different levels of integration, we can also have different modes of weight sharing when there are parallel network layers corresponding to different pulse sequences, as in the mid-level and end-level architectures.  We considered three major cases: (1) fully independent, (2) fully shared, and (3) L2-tied.  \textit{Fully independent weights} have parallel layers that are initialized randomly and receive different gradients.  \textit{Fully shared weights} (also referred to as fully siamese), are initialized identically and receive identical gradients.  \textit{L2-tied weights} are initialized identically and allowed to receive different gradients, subject to an L2 loss on pairs of parallel weights to penalize diverging weights.  These loss functions are represented in a high-level mathematical fashion in equations 1-3.  In both the input- and end-level integration schemas, we only have one basic network that we will call $F$ with corresponding learned parameters $\theta$.  However, for the mid-level integration network, we will call the convolution blocks before integration $F_e$ (for early) and the blocks after integration $F_l$ (for late), with corresponding parameters $\theta_e$ and $\theta_l$.  Cases where weights are allowed to diverge (receive differing gradients) or forced to be identical are denoted by separating a single $\theta$ value into $\{\textcolor{OliveGreen}{\theta_1}, \textcolor{Blue}{\theta_2}, \textcolor{Dandelion}{\theta_3}, \textcolor{Plum}{\theta_4}\}$.  Differing colors match those seen in figure \ref{fig:integrationLevels}.

\begin{center}

\begin{equation}\tag{1}\label{eq:1}
\begin{split}
\hat{Y} &= F(\textcolor{OliveGreen}{X_1}, \textcolor{Blue}{X_2}, \textcolor{Dandelion}{X_3},\textcolor{Plum}{X_4}; \theta) \\
\mathcal{L} &= \text{CE}(Y, \hat{Y})
\end{split}
\end{equation}

\rule{3in}{1pt} \vspace{-4mm}

\begin{align*}
\hat{Y} &= F_l( F_e (\textcolor{OliveGreen}{X_1};\textcolor{OliveGreen}{\theta_1}), F_e (\textcolor{Blue}{X_2};\textcolor{Blue}{\theta_2}),F_e (\textcolor{Dandelion}{X_3};\textcolor{Dandelion}{\theta_3}),F_e (\textcolor{Plum}{X_4};\textcolor{Plum}{\theta_4}); \theta_l) \\
\mathcal{L} &= \text{CE}(Y, \hat{Y})\tag{2a}\label{eq:2a}
\end{align*}

\rule{3in}{0.5pt} \vspace{-4mm}

\begin{align*}
\hat{Y} &= F_l( F_e (\textcolor{OliveGreen}{X_1};\theta_e), F_e (\textcolor{Blue}{X_2};\theta_e),F_e (\textcolor{Dandelion}{X_3};\theta_e),F_e (\textcolor{Plum}{X_4};\theta_e); \theta_l) \\
\mathcal{L} &= \text{CE}(Y, \hat{Y})\tag{2b}\label{eq:2b}
\end{align*}

\rule{3in}{0.5pt} \vspace{-4mm}

\begin{align*}
\hat{Y} &= F_l( F_e (\textcolor{OliveGreen}{X_1};\textcolor{OliveGreen}{\theta_1}), F_e (\textcolor{Blue}{X_2};\textcolor{Blue}{\theta_2}),F_e (\textcolor{Dandelion}{X_3};\textcolor{Dandelion}{\theta_3}),F_e (\textcolor{Plum}{X_4};\textcolor{Plum}{\theta_4}); \theta_l) \\
\mathcal{L} &= \text{CE}(Y, \hat{Y}) + \sum_{i=1}^4 L_2 (\theta_i, \bar{\theta})\tag{2c}\label{eq:2c}
\end{align*}

\rule{3in}{1pt} \vspace{-4mm}

\begin{align*}
\hat{Y} &= F(\textcolor{OliveGreen}{X_1},\textcolor{OliveGreen}{\theta_1}) + F(\textcolor{Blue}{X_2},\textcolor{Blue}{\theta_2}) + F(\textcolor{Dandelion}{X_3},\textcolor{Dandelion}{\theta_3}) + F(\textcolor{Plum}{X_4},\textcolor{Plum}{\theta_4}) \\
\mathcal{L} &= \text{CE}(Y, \hat{Y})\tag{3a}\label{eq:3a}
\end{align*}

\rule{3in}{0.5pt} \vspace{-4mm}

\begin{align*}
\hat{Y} &= F(\textcolor{OliveGreen}{X_1},\theta) + F(\textcolor{Blue}{X_2},\theta) + F(\textcolor{Dandelion}{X_3},\theta) + F(\textcolor{Plum}{X_4},\theta) \\
\mathcal{L} &= \text{CE}(Y, \hat{Y})\tag{3b}\label{eq:3b}
\end{align*}

\rule{3in}{0.5pt} \vspace{-4mm}

\begin{align*}
\hat{Y} &= F(\textcolor{OliveGreen}{X_1},\textcolor{OliveGreen}{\theta_1}) + F(\textcolor{Blue}{X_2},\textcolor{Blue}{\theta_2}) + F(\textcolor{Dandelion}{X_3},\textcolor{Dandelion}{\theta_3}) + F(\textcolor{Plum}{X_4},\textcolor{Plum}{\theta_4}) \\
\mathcal{L} &= \text{CE}(Y, \hat{Y})+ \sum_{i=1}^4 L_2 (\theta_i, \bar{\theta})\tag{3c}\label{eq:3c}
\end{align*}

\end{center}

\subsection{Special Case: Combining Pre- and Post-Contrast}

We have created a special architecture for integrating pre- and post-contrast images.  In addition to testing all combinations of integration levels and weight-sharing modes, we include an additional mid-level integration scheme.  Rather than concatenating the pre- and post-contrast layers, we combine them and subtract them in the feature-space.  Our reasoning for this is that the features computed from the pre-contrast images and the features from the post-contrast images should only differ locally where the imaging differs, which should correspond to lesions.  This choice, which is informed by domain knowledge, may be favorable for the bias-variance tradeoff.  Note that this approach makes sense only if the weights in the pre- and post-contrast branches are fully shared.

\subsection{Integration-Level Dropout Layer}

A key problem is that a network trained with a given structure (i.e., given preset input channels) is not robust to any missing information during inference.  To address this limitation, we propose stochastically zeroing out random pulse sequences during training, an idea inspired by the channel-wise dropout commonly used when training CNNs.  By training on inputs with missing pulse sequences, such networks should be robust to similar inputs during deployment.

\begin{figure}[htb]
	\centering
	\includegraphics[width=0.45\textwidth]{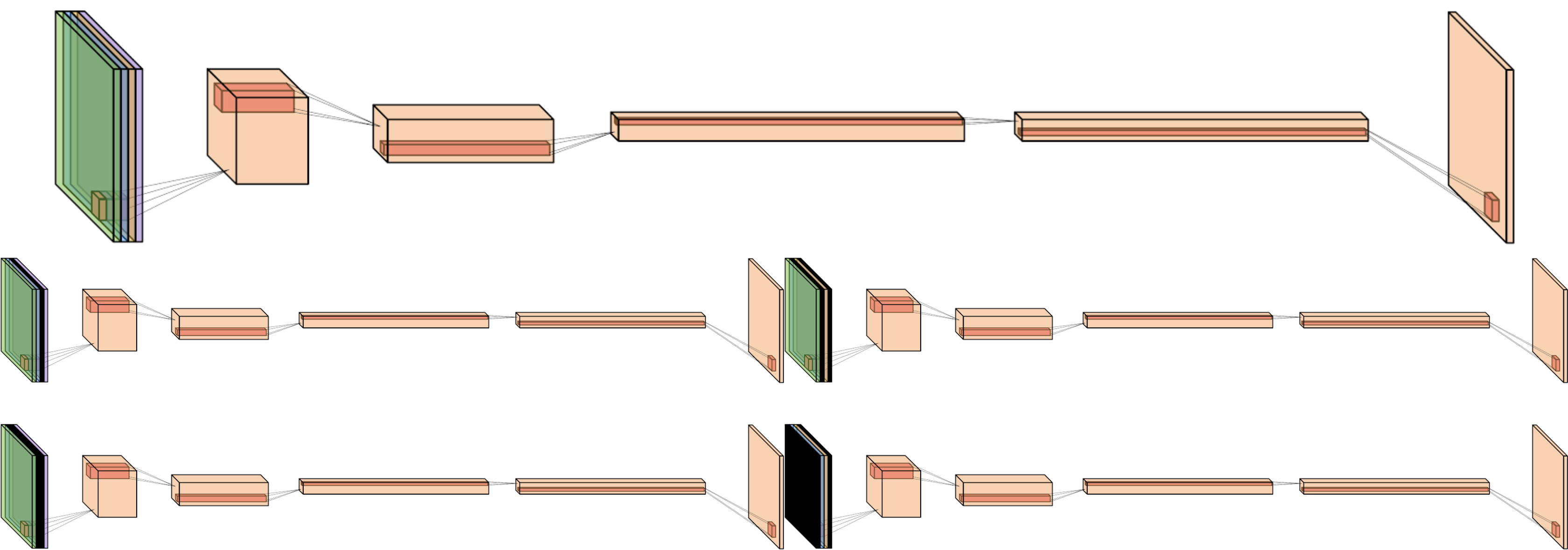}
	\caption{\textbf{Integration Level Dropout for Input-Level Integration}  \small{We take the standard input-level integration architecture (reproduced from Figure \ref{fig:integrationLevels}a, top) and show four potential “dropped out” versions that we might encounter during training (bottom).  From our original four input-concatenated pulse sequences, we randomly drop out 0-3 input channels (shown in black).  Thus, the network can encounter missing pulse sequences during training.}}
	\label{fig:dropout}
\end{figure}

An important consideration is the weighting function.  Because dropping out neurons with probability $p$ during training but having them fully firing during inference creates a difference in the total sum of neuron activations, we can either upweight the dropout layer’s neurons by a factor of $\frac1{1-p}$ during training or downweight the dropout layer’s neurons by a factor of $1-p$ during inference.  We upweight our input dropout layer by a factor of $\frac1{1-p}$ where $p$ is the actual proportion of dropped out pulse sequences, in both training and inference.

An important note is that dropout cannot be done on a purely statistical basis.  Normally, each channel has a certain probability of being dropped.  However, as we use a 2.5D network structure, we must take care to drop out all of the z-slices of a certain pulse sequence or none at all.  We must also make sure to never drop out all four pulse sequences during training.  This situation would result in the network receiving an input tensor of all 0s, which would be intractable and lead to unstable training.

\subsection{Visualizing Input Gradient Accumulation}\label{sec:saliency}

A standard method for visualizing networks is the saliency map, which in its most basic form is the gradient of the loss with respect to the input image:

\begin{equation}\tag{4}\label{eq:4}
\text{Saliency Image} = \frac{\partial L}{\partial \text{Img}}
\end{equation}

From equation 4, we construct separate quantities that we term the \emph{saliency aggregate} and the \emph{cumulative saliency aggregate}.  The saliency aggregate is the sum of the absolute value of the saliency image, which approximates the sensitivity of the loss with respect to the input image.  The cumulative saliency aggregate is the iteration-based cumulative sum of the saliency aggregate during training. These metrics allow us to quantify how the network learns over the course of training.

\begin{equation}\tag{5a}\label{eq:5a}
\text{Saliency Aggregate} = \text{Sum} \left( \left| \frac{\partial L}{\partial \text{Img}} \right| \right)
\end{equation}

\begin{equation}\tag{5b}\label{eq:5b}
\text{Cumulative Saliency Aggregate} = \sum_{\text{iter} = 0}^{\text{now}} \text{Sum} \left( \left| \frac{\partial L}{\partial \text{Img}} \right| \right)_{\text{iter}}
\end{equation}

\begin{figure*}[htb]
	\centering
	\includegraphics[width=0.95\textwidth]{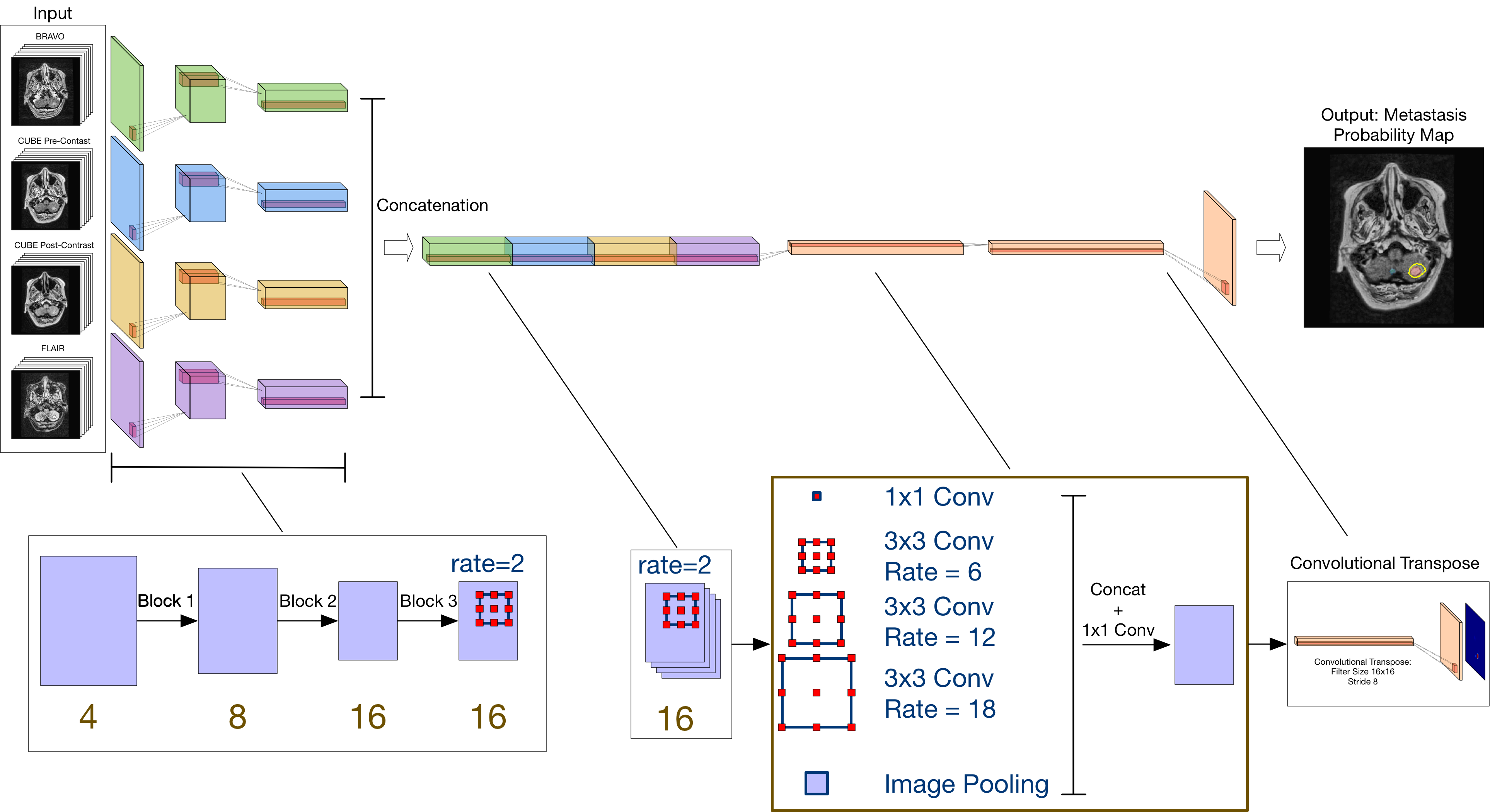}
	\caption{\textbf{Deep Learning Pipeline}  \small{Our main pipeline is simple and is built largely on the DeepLabv3 architecture \cite{chen2017rethinking}.  For our input-level integration models, we will concatenate all of our input pulse sequence slices together.  For our mid-level integration models, we concatenate the features immediately before the ASPP layer, as shown in this figure.  Finally, for end-level integration, we will pass through each pulse sequence through an independent DeepLabv3 model and average the output probability maps.}}
	\label{fig:pipeline}
\end{figure*}

\subsection{Implementation Details}\label{sec:implementationdetails}

In our paper, we use the DeepLabv3 \cite{chen2017rethinking} architecture as our segmentation baseline.  The architecture is well known for its ability to incorporate an extremely large field of context, most likely due to its atrous (dilated) convolutions.  Though a potential weakness of the DeepLabv3 is worse performance on thinner structures (such as the legs of a chair or a light pole), this is not a significant concern for metastasis segmentation, as most lesions are spherical.  For mid-level integration architectures, we integrate the modalities after the fourth main residual block, right before the Atrous Spatial Pyramid Pooling (ASPP) layer.

Our input pulse MR’s have all been resized to a 256x256 image.  The only preprocessing that has been done is adaptive histogram equalization on each slice of the data.  This provides two main benefits: (1) each pixel value for every scan is in the same value-space and (2) contrast is enhanced.  We believe that at the cost of minimal artifacts, adaptive histogram equalization is a quick and dirty way to fix domain shift errors to a first approximation.  For our 2.5D, we use 5 slices.  Thus, for our input-level integration networks, we will have an input tensor with 20 channels, 5 slices each for 4 pulse sequences.

To run dropout experiments, we create a custom data loader transform in PyTorch, which takes input data of four pulse sequences and randomly drops out each pulse sequence with a probability of 25\%.  In the case that all four pulse sequences are dropped out (probability ~0.4\%), no pulse sequence is dropped.  The table below shows the probability distribution for the number of pulse sequences dropped out.

\begin{table}[htb]
\begin{center}
\begin{tabular}{C{2.4cm}|C{0.7cm}C{0.7cm}C{0.7cm}C{0.7cm}C{0.7cm}}
\# Seq's Dropped Out&0&1&2&3&4\\
\midrule
Probability&32\%&42\%&21\%&4.7\%&0\%
\end{tabular}
\end{center}
\end{table}

Furthermore, our main experiments for missing pulse sequences rely on censoring out the BRAVO Post-Contrast sequence.  In our training, there is an 11\% chance of seeing an input image with only the BRAVO sequence dropped out.  

All code was written in PyTorch.  Both training and inference were done on two NVIDIA 1080Ti GPUs.  During training, we over-sample z-slices with any portion of ground truth annotated lesions by a factor of 10x compared to frames without.  Training was done for 10 epochs from random initialization on a dataset of 100 patients, taking about 20 hours.  The forward pass inference time is approximately 150ms per z-slice, or about 30 seconds for a single patient.  This becomes about 5 seconds per z-slice on a CPU (Intel i7-8700k), which translates to a 30-minute patient inference time.

\subsection{Metrics for Detection and Segmentation}

In our work, we treat the segmentation network as both a detection and a segmentation network.  To evaluate detection performance, we report the mean average precision (mAP) value or area under the precision-recall (PR) curve.  Generating a PR curve requires (1) a standard to decide when a prediction is a true positive and (2) a probability per predicted lesion.  We create predicted lesions by binarizing the complete 3D segmentation probability volume, using an empirically determined probability threshold of 10\%.  Thus, any 3D connected component of predicted voxels with probability greater than 10\% is a single predicted lesion.  If the center of mass of the 3D connected component is within 1mm of a ground truth annotation, we call that predicted lesion a true positive.  (Given that lesions are small, this metric is conservative.)  The predicted probability of the lesion will be the average probability of all of the voxels in that region with respect to the original probability volume.  Thus, by definition, the minimum predicted probability for any lesion with this method will be 10\% or 0.1.  With the constructed PR curve, we report the mAP score and the maximum sensitivity.

To quantify segmentation performance, we report DICE scores for our true positive segmentations at the maximum sensitivity.  In other words, we take our binary segmentation map (with probability threshold 10\%) and report the DICE scores for all predicted binary lesions with center of mass within 1mm of the expert-annotated lesions.

\section{Results}

\subsection{Stanford Metastasis Dataset Experiments}

\begin{table}[htb]
 \caption{\textbf{Pulse Sequence Integration}}\label{table:pulseSeqInt}
\begin{center}
\begin{tabular}{C{1.4cm}|C{1.8cm}|C{1.2cm}C{0.8cm}C{0.8cm}}
\toprule[1.5pt]
Integration Level & Weight Sharing & mAP & Max Sens. & TP DICE \\
\midrule
\midrule
Input & - & \underline{46} (44,47) & 80 & 72 \\
Mid & Fully Shared & 44 (42,45) & 82 & \underline{74} \\
Mid & Fully Ind. & 39 (37,41) & 54 & 70 \\
Mid & L2 Tied & 44 (42,45)& \underline{83} & 73 \\
End & Fully Shared & 35 (33,37)& 52 & 70 \\
End & Fully Ind. & 28 (25,30)& 47 & 62 \\
End & L2 Tied & 34 (32,36) & 53 & 71 \\
\midrule
\multicolumn{5}{l}{Below are values for the pre- and post-contrast subtraction.} \\
\midrule
Mid & Fully Shared & \textbf{48} (46,49) & \textbf{86} & \textbf{75}\\
\midrule
\midrule
\end{tabular}
\end{center}
\end{table}

Table \ref{table:pulseSeqInt} shows results from the different pulse sequence integration and weight sharing architectures.  The best values are bolded, while the second-best values are underlined.  In general, the best performing network is the pre-/post-contrast subtraction mid-level integration network. The next best networks are the input-level integration network and the fully shared and L2-tied mid-level integration networks.

\begin{table}[htb]
 \caption{\textbf{Pulse Sequence Int. with Dropout Training}}\label{table:pulseSeqIntDO}
\begin{center}
\begin{tabular}{C{1.3cm}|C{1.5cm}|C{1.4cm}C{1.4cm}C{1.3cm}}
\toprule[1.5pt]
Integration Level & Weight Sharing & mAP & Max Sens. & TP DICE \\
\midrule
\midrule
Input & - & \textbf{45} (-2.2\%) & 80 (0.0\%) & 72 (0.0\%) \\
Mid & Fully Shared & 43 (-2.3\%) & \textbf{83} (+1.2\%) & \textbf{74} (0.0\%) \\
Mid & Fully Ind. & 34 (-12.8\%) & 44 (-18.5\%) & 71 (+1.4\%) \\
Mid & L2 Tied & \textbf{45} (+2.3\%) & \textbf{83} (0.0\%) & \underline{73} (0.0\%) \\
End & Fully Shared & 35 (0.0\%) & 54 (+3.8\%) & 70 (0.0\%) \\
End & Fully Ind. & 24 (-14.3\%) & 41 (-12.8\%) & 58 (-6.5\%) \\
End & L2 Tied & 34 (0.0\%) & 50 (-5.7\%) & 71 (0.0\%) \\
\midrule
\midrule
\end{tabular}
\end{center}
\end{table}

Table \ref{table:pulseSeqIntDO} shows the results of training the networks with the input dropout schema.  There is a significant decrease in performance in the two networks with fully independent weight sharing.  In addition, we see some network degradation for the L2-tied end-level integration network.  However, other networks, including the high-performing networks from Table \ref{table:pulseSeqInt} (input-level integration, fully shared mid-level, and L2-tied mid-level), show approximately equivalent performance.

The pre-/post-contrast subtraction network was not trained with input dropout, as dropping out either pre- or post-contrast images alone disrupted the core structure of the network.

\subsection{Input Pulse Sequence Censorship}

\begin{table}[htb]
 \caption{\textbf{Non-Dropout Model on BRAVO-Censored Data}}\label{table:noDO}
\begin{center}
\begin{tabular}{C{1.4cm}|C{1.8cm}|C{1.2cm}C{0.8cm}C{0.8cm}}
\toprule[1.5pt]
Integration Level & Weight Sharing & mAP & Max Sens. & TP DICE \\
\midrule
\midrule
Input & - & 0 (0,0) & 0 & - \\
Mid & Fully Shared & 0 (0,0) & 0 & - \\
Mid & Fully Ind. & 0 (0,0)& 0 & - \\
Mid & L2 Tied & 0 (0,0)& 0 & - \\
End & Fully Shared & 0 (0,0)& 0 & - \\
End & Fully Ind. & 0 (0,0)& 0 & - \\
End & L2 Tied & 0 (0,0)& 0 & - \\
\midrule
\multicolumn{5}{l}{Below is an input-level int. model trained without BRAVO.} \\
\midrule
Input & - & \textbf{40} (38,41) & \textbf{65} & \textbf{71}\\
\midrule
\midrule
\end{tabular}
\end{center}
\end{table}

Table \ref{table:noDO} highlights the need for a network robust to missing pulse sequences.  When censoring out the BRAVO pulse sequence during inference on a network trained with all four pulse sequences, the network’s predictive power is lost.  One simple solution is to train a network for the set of pulse sequences without the BRAVO, in anticipation of this case.  As shown in the last row of Table \ref{table:noDO}, this network has good performance, albeit degraded, in inference on data without the BRAVO pulse sequence.

\begin{table}[htb]
 \caption{\textbf{Dropout Model on BRAVO-Censored Data}}\label{table:DO}
\begin{center}
\begin{tabular}{C{1.4cm}|C{1.8cm}|C{1.2cm}C{0.8cm}C{0.8cm}}
\toprule[1.5pt]
Integration Level & Weight Sharing & mAP & Max Sens. & TP DICE \\
\midrule
\midrule
Input & - & 37 (35,39) & 62 & \underline{72} \\
Mid & Fully Shared & \textbf{40} (38,41)& \underline{64} & 70 \\
Mid & Fully Ind. & 34 (32,36)& 60 & \textbf{73} \\
Mid & L2 Tied & \underline{38} (36,40)& \textbf{65} & 71 \\
End & Fully Shared & 30 (27,32)& 55 & 70 \\
End & Fully Ind. & 15 (12,18)& 45 & 63 \\
End & L2 Tied & 28 (25,30)& 57 & \underline{72} \\
\midrule
\midrule
\end{tabular}
\end{center}
\end{table}

Table \ref{table:DO} shows the results of inference on a BRAVO-censored dataset using networks trained with the input dropout technique, all of which show restored performance.  Though none of the networks matches the level of the model trained on the exact subset of inference pulse sequences (last row of Table \ref{table:noDO}), performance is approximately equivalent with the fully shared mid-level integration network trained with dropout. This demonstrates the enhanced robustness to missing pulse sequences conferred by the input dropout layer.

\begin{table}[htb]
 \caption{\textbf{Input-Level Dropout Model on Different Combinations of Input Pulse Sequences}}\label{table:combos}
\begin{center}
\begin{tabular}{C{1.0cm}|C{0.8cm}|C{0.8cm}|C{0.9cm}|C{1.2cm}C{0.6cm}C{0.7cm}}
\toprule[1.5pt]
BRAVO Post-C & CUBE Pre-C & CUBE Post-C & FLAIR & mAP & Max Sens. & TP DICE \\
\midrule
\midrule
\cmark & \cmark & \cmark & \cmark & \textbf{45} (43,46)& \textbf{80} & \textbf{72} \\
\xmark & \cmark & \cmark & \cmark & 37 (35,39)& 62 & \textbf{72} \\
\cmark & \xmark & \cmark & \cmark & 38 (36,40)& 67 & 71 \\
\cmark & \cmark & \xmark & \cmark & 42 (40,43)& 72 & 70 \\
\cmark & \cmark & \cmark & \xmark & \underline{44} (42,45)& \underline{79} & \textbf{72} \\
\midrule
\midrule
\end{tabular}
\end{center}
\end{table}

Having trained networks that can utilize any subset of the four predefined pulse sequences, the relative utility of each modality can be assessed by censoring a particular pulse sequence and using that test set in the input-level dropout model. Table \ref{table:combos} compares the sensitivity of the network when each pulse sequence is omitted in turn.  Unsurprisingly, the network performs best when all of the pulse sequences are given during inference.  However, it is most sensitive to the loss of the post-contrast BRAVO and the pre-contrast CUBE, followed by loss of the post-contrast CUBE.  The network is minimally sensitive to the absence of the T2-weighted FLAIR.

\subsection{Evaluation on Oslo Dataset}

To evaluate the generalizability of the trained network, we evaluated its performance on data from a different center (Oslo).  In table \ref{table:Oslo} below, we see a comparison of the dropout-based networks and the exact-subset network.  The network trained on all four pulse sequences without the dropout had complete failure, the same as when censoring out BRAVO data on the Stanford test set.  All networks below are the same networks shown in previous sections, i.e. trained only on the 100-patient Stanford training set.  These values show not only that the networks generalized well to the Oslo data but also that the exact-subset network and the dropout-trained networks showed comparable performance.

It should be noted here that the bottom row represents two different models.  Of the 67-patient Oslo cohort, 65 patients had 3 pulse sequences: (1) CUBE pre-contrast, (2) CUBE post-contrast, and (3) T2w FLAIR.  However, 2 patients only had 2 pulse sequences: (1) CUBE post-contrast and (2) T2w FLAIR.  Thus, we used a network trained on the exact subset of three pulse sequences on the 65 patients and another network trained on the exact subset of two pulse sequences for the remaining two patients.  For the dropout model, we only used one model per row but did change the weighting factor between the two groups of patients, as discussed in Section \ref{sec:implementationdetails}.

\begin{table}[htb]
 \caption{\textbf{Testing on Oslo Data (no BRAVO Scans)}}\label{table:Oslo}
\begin{center}
\begin{tabular}{C{1.4cm}|C{1.8cm}|C{1.2cm}C{0.8cm}C{0.8cm}}
\toprule[1.5pt]
Integration Level & Weight Sharing & mAP & Max Sens. & TP DICE \\
\midrule
\midrule
Input & - & \textbf{68} (66,69) & \textbf{92} & \textbf{85} \\
Mid & Fully Shared & 67 (65,69)& 90 & 83 \\
Mid & Fully Ind. & 61 (58,63)& 75 & 82 \\
Mid & L2 Tied & \textbf{68} (66,69)& \textbf{92} & \underline{84} \\
End & Fully Shared & 59 (57,60)& 70 & 80 \\
End & Fully Ind. & 36 (33,38)& 45 & 72 \\
End & L2 Tied & 55 (53,57)& 67 & 81 \\
\midrule
\multicolumn{5}{l}{Below is an input-level int. model trained on without BRAVO.} \\
\midrule
Input & - & 67 (65,69) & \textbf{92} & \underline{84}\\
\midrule
\midrule
\end{tabular}
\end{center}
\end{table}

A representative image comparing the baseline input-level network, the same network trained with dropout, and the same network trained without dropout on the exact subset data (BRAVO-censored) has been shown in figure \ref{fig:visualization}.  Worth noting in figure \ref{fig:visualization} is the fact that our baseline model will have complete failure on the Oslo patient, since it did not receive all four pulse sequences it was trained with.  However, both the dropout model and the model trained without BRAVO perform well on the Oslo Patient.

\subsection{Visualization with Saliency Map Cumulative Sum}

\begin{figure}[htb]
	\centering
	\begin{subfigure}[h]{0.22\textwidth}
		\includegraphics[width=\textwidth]{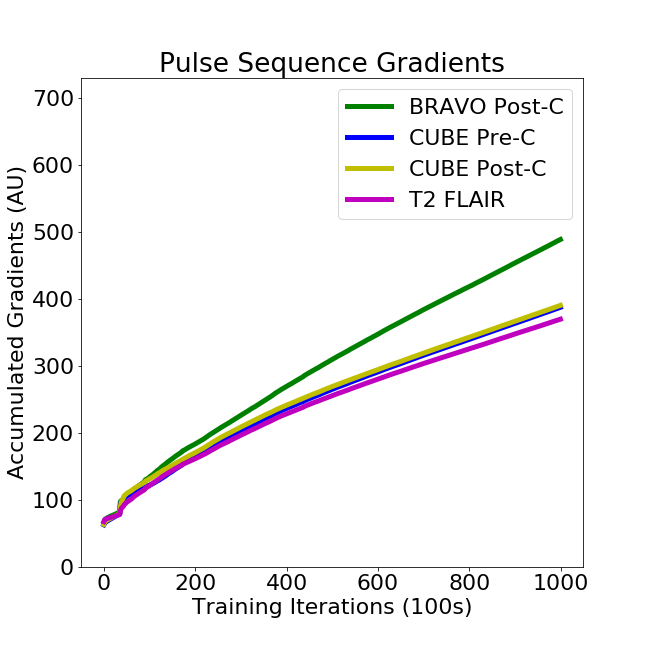}
		\caption{Pulse Sequence Gradients}
	\end{subfigure}
	~
	\begin{subfigure}[h]{0.22\textwidth}
		\includegraphics[width=\textwidth]{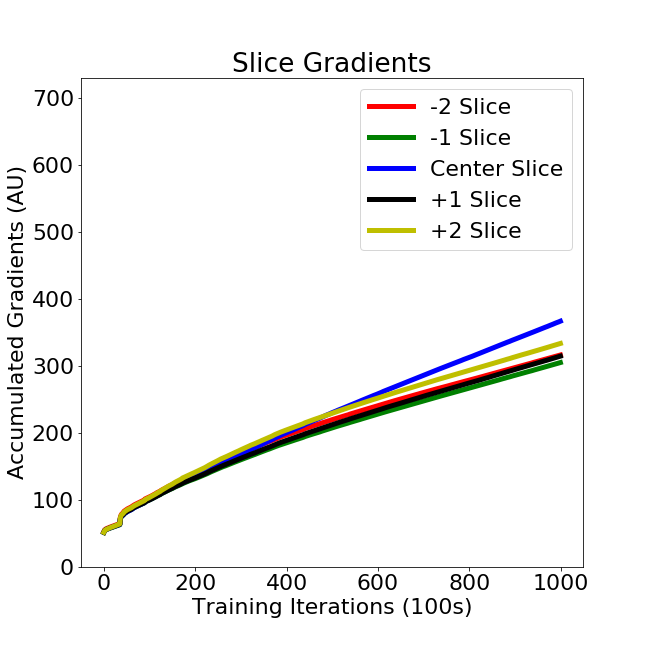}
		\caption{Slice Gradients}
	\end{subfigure}
	\caption{\textbf{Cumulative Aggregate Saliencies for Network Trained without Dropout.}}
	\label{fig:nodo1}
\end{figure}

\begin{figure}[htb]
	\centering
	\begin{subfigure}[h]{0.22\textwidth}
		\includegraphics[width=\textwidth]{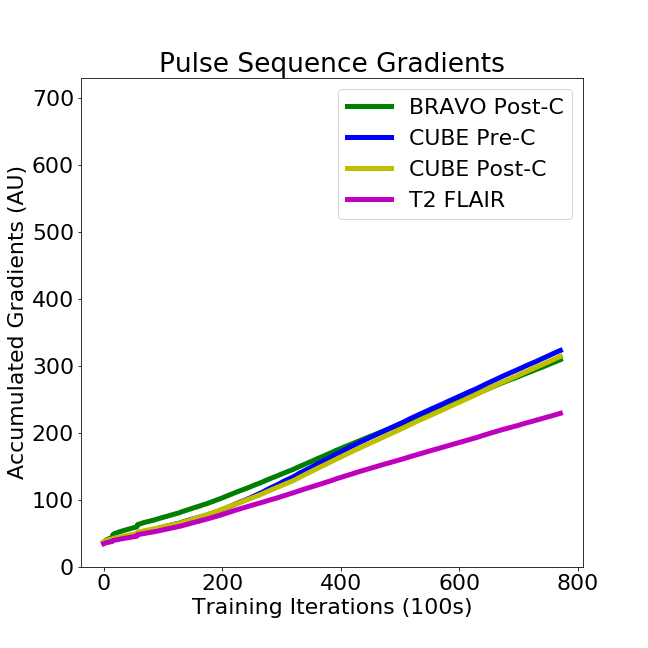}
		\caption{Pulse Sequence Gradients}
	\end{subfigure}
	~
	\begin{subfigure}[h]{0.22\textwidth}
		\includegraphics[width=\textwidth]{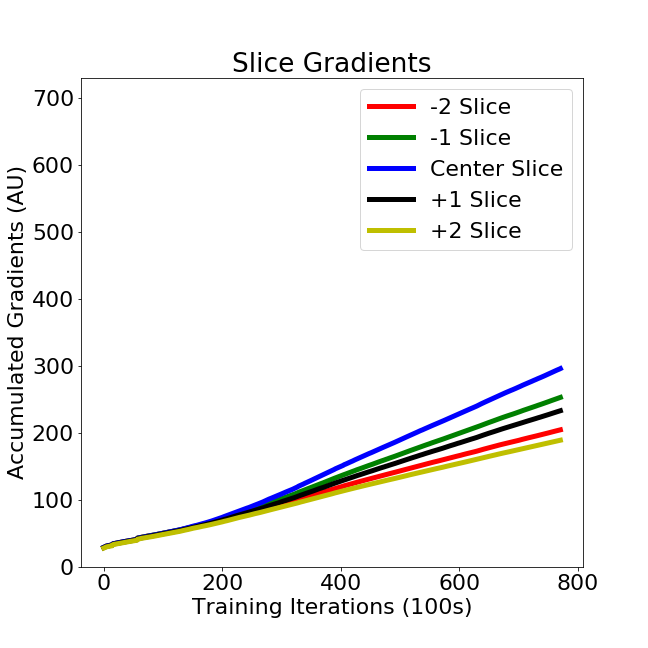}
		\caption{Slice Gradients}
	\end{subfigure}
	\caption{\textbf{Cumulative Aggregate Saliencies for Network Trained with Dropout.}}
	\label{fig:do1}
\end{figure}

Below, we investigate the cumulative aggregate saliency (Equation \ref{eq:5b}) for our input-level integration networks trained with and without dropout.  In Figures \ref{fig:nodo1} and \ref{fig:do1}, we look at a single input-level integration network (corresponding to the input-level integration networks shown in Tables 1-6).  On the left, we see the cumulative aggregate saliencies separated by input pulse sequences.  On the right, we see the cumulative aggregate saliencies separated by z-slice position.

\begin{figure}[htb]
	\centering
	\includegraphics[width=0.45\textwidth]{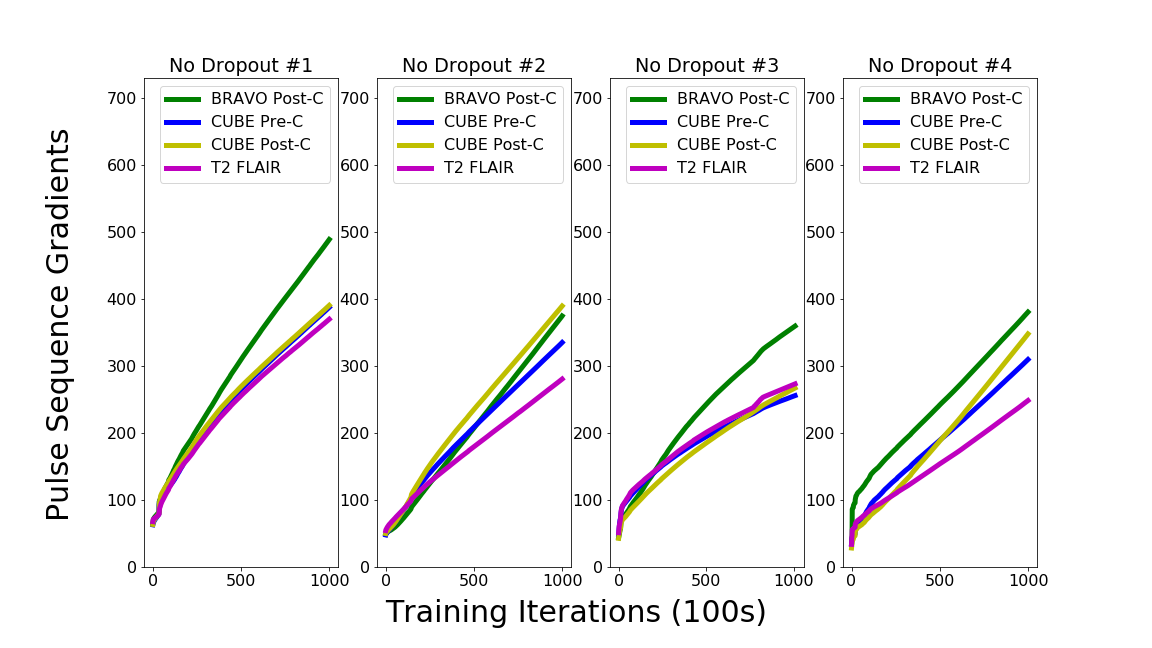}
	\caption{\textbf{Four Stochastically Different Training Runs of No-Dropout Networks.}}
	\label{fig:nodotot}
\end{figure}

\begin{figure}[htb]
	\centering
	\includegraphics[width=0.45\textwidth]{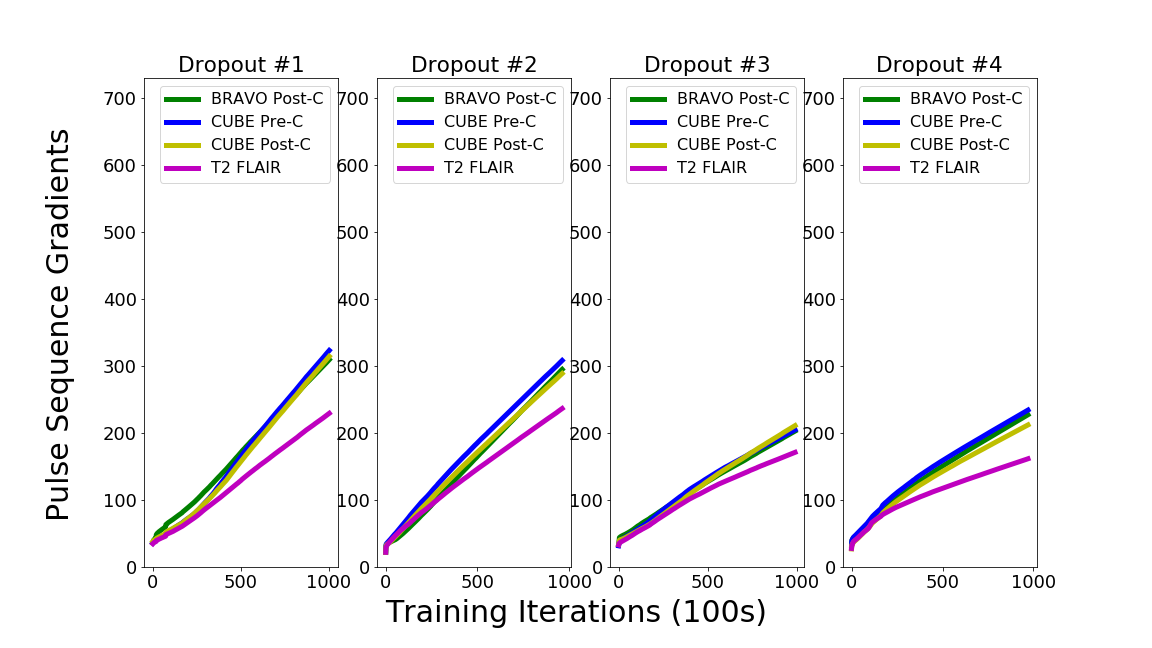}
	\caption{\textbf{Four Stochastically Different Training Runs of Dropout Networks.}}
	\label{fig:dotot}
\end{figure}

We also trained three additional input-level networks each with and without the dropout method.  The networks were trained identically apart from random parameter initializations and different mini-batch orderings.  These networks, along with the original network, have their pulse sequence saliencies plotted in Figures \ref{fig:nodotot} and \ref{fig:dotot}.  These results demonstrate that the dropout method allows for a more consistent training schedule, learning at approximately equal rates from all available pulse sequences besides FLAIR.  Additional quantitative results are not shown for these three networks, but they are comparable to those of the original network.

\section{Discussion}

\subsection{Pulse Sequence Integration and the Bias-Variance Tradeoff}

In general, the best networks have the most shared parameters or most regularization (Table \ref{table:pulseSeqInt}).  This leads us to believe that the small dataset (100 patients) favors constraints on networks and precludes identification of the optimal architecture.  It is clear that the total compute of the input-level integration network is the lowest, as it lacks replicated layers.  Since mid-level and end-level integration show no clear benefit, we advise using the input-level integration network in the small data regime.

Intriguingly, the best network is a mid-level integration network where the pre- and post-contrast CUBE features are subtracted at the integration level rather than concatenated.  We believe that this result fits with the bias-variance tradeoff in the small data regime.  By constructing a high-bias model that only the difference of pre- and post-contrast features should be important, we constrain our model in a way that minimizes variance error.  However, we argue that the assumptions behind subtracting the pre- and post-contrast features are well founded, adding minimal bias error to the model.



\subsection{Dealing with Missing Pulse Sequences}

As shown in Table \ref{table:noDO}, missing pulse sequences destroyed the performance of our trained models, motivating us to develop a single network robust to missing pulse sequences at inference time.  Such a network would be essential for real-world applications, as the types of data taken may vary over time or between institutions.  Training with integration-level dropout restored network performance, as shown in Table \ref{table:DO}.  We compared the performance of these models to that of a network trained on the exact subset of provided data without the dropout method, given in the last row of Table \ref{table:noDO}.  Though this exact-subset network performs best overall, the fully-shared and L2-tied mid-level integration dropout models closely match its performance.  We can also see from Table \ref{table:pulseSeqInt} that introducing the dropout method does not degrade performance when we are given all pulse sequences.

One important consideration is the difference in model selection between the exact-subset and dropout networks.  Every 1000 iterations, we tested our model on a small validation set (n=6).  If the model performed better than previously, we saved that state as the new best model, overwriting the previous best.  The exact-subset network was validated on a set that had the exact subset, while the dropout models were validated on inputs with all four pulse sequences available.  Thus, the dropout models were biased for the best performance given no missing pulse sequences.  We additionally trained an input-level integration network with dropout but with the validation set having the BRAVO scan censored out.  This created a model that performed slightly better than the exact subset network (mAP=42, max sensitivity=69, TP DICE=72).  However, performance was slightly reduced when testing on the test set without any censorship (mAP=44, max sensitivity=75, TP DICE=72).

\subsection{Input Dropout Network to Assay Pulse Sequence Importance}

Having trained a flexible network that could use any subset of pulse sequences, we sought to leverage this feature to probe changes in performance given different input pulse sequence combinations, as shown in Table \ref{table:combos}.  By censoring each of the input pulse sequences individually, we found the largest degradations in performance when censoring the post-contrast BRAVO or the pre-contrast CUBE.  Since the post-contrast BRAVO was used to create the annotations and all other sequences were registered to the BRAVO, this scan would be free from registration errors and would have the highest spatial correlation to expert annotations.  We also think that the pre-contrast CUBE serves as a point of comparison with respect to the post-contrast scans.  On the other hand, the post-contrast CUBE is less important, likely because the BRAVO sequence, which is also a post-contrast image, captures similar information.  Finally, missing the FLAIR has very little impact on the network’s performance.  The FLAIR is often used in clinical practice to indicate edema and other fluid as a larger marker for localizing lesions, but our annotations center on the core metastatic lesions, not edema.  Thus, the FLAIR signal does not correlate with the primary annotations.

Additionally, we can look at the cumulative aggregate saliency for the input image, as described in Section \ref{sec:saliency}.  Figure \ref{fig:nodo1} shows the input image cumulative aggregate saliency when separated by (a) pulse sequence and (b) z-slice location.  Looking at the z-slice location, we see very predictable behavior, with the aggregate saliency being strongest when centered on the location of the annotation, followed by the +1/-1 slices, and then finally the +2/-2 slices.  Repeating this analysis for the pulse sequences of a no-dropout model shows that the BRAVO post-contrast images have the largest gradients.  However, the behavior of cumulative aggregate saliency can change between training runs with otherwise identical hyperparameters (Figure \ref{fig:nodotot}), although the BRAVO scan often eventually shows the largest gradients.  Comparing Figures \ref{fig:do1} and \ref{fig:dotot} (which visualize the models trained with input-level dropout), we observe that the dropout training creates a regular training pattern over time.  When we look at the saliencies for different dropout runs, we notice the same pattern: all sequences other than the FLAIR train at approximately equal rates, while the FLAIR is always found to be the least correlated with the annotations.  We believe that since our expert annotations did not segment edema portions of the metastasis, there was minimal correlation between the FLAIR images and our segmentation targets.  In addition, adding dropout regularization creates a more consistent training behavior.  This could be meaningful both scientifically and practically to create a consistent set of networks.  However, this property could be detrimental if stochasticity in the networks is desired, such as in ensembling many networks.

\subsection{Robustness to Inference of Multi-center Data}

By testing our model on the Oslo data, we provide a real-world use case for training our models with input-level dropout, as these data do not have the same pulse sequences as that of our training data from Stanford.  Standard practice involves training a separate model for every subset of pulse sequences, an exponentially difficult task with respect to number of pulse sequences.  (For n pulse sequences, we would have to train (2n-1) models.)  However, in table \ref{table:Oslo}, we show comparable performance on the Oslo data between the models trained with and without dropout.

We also notice a substantial boost in performance on the Oslo test set compared to the Stanford test set.  We attribute this to the fact that the Oslo data had, on average, much larger metastasis lesions, which should be more easily detected.  The Oslo data was also taken at a lower slice thickness, which gave more resolution for prediction.

\section{Conclusion}

We set out to investigate best practices for integrating different pulse sequences of MR data using metastasis segmentation as a model task.  We found that within the small data regime (n=100), models that had lower variance error performed better.  Our best performing model was one that encoded in a subtraction between the learned-features from the pre- and post-contrast images.  This leads us to believe that leveraging the clinical relationship between pulse sequences and mathematically encoding this knowledge in our network architecture could potentially boost network performance without requiring more data.

We also created and tested using an input dropout layer to train a single network that was robust to receiving any subset of the input pulse sequences (excluding the trivial case of an empty subset).  With this method, the network not only preserves performance when given the full set of pulse sequences but also rivals the results of a model trained on the exact subset of pulse sequences when tested on that subset of pulse sequences.  Using this dropout network, we were able to show that a single network could be trained on data from one institution (Stanford Hospital) and generalize to test data from another institution (Oslo University Hospital).

Finally, we ran network interpretation methods to better understand how our new dropout network performs in different circumstances.  By running different subsets of pulse sequences through our network, we revealed which pulse sequences were most important in the decision-making process of our network.  By visualizing the accumulated image gradients throughout training, we were able to show that the dropout network not only learned more evenly from the input pulse sequences but also gave more consistent training schedules.  These results provide new insight on how CNNs can be most effectively applied to the multi-modal problem of medical images.

\section*{Acknowledgment}

We'd like to thank both Stanford Hospital and Oslo University Hospital for providing the data needed to complete this study.  We acknowledge the T15 LM 007033 NLM Training grant in funding this project.  This work was also supported in part by grants from the National Cancer Institute, National Institutes of Health, U01CA142555, 1U01CA190214, 1U01CA187947 and U01CA242879.

\ifCLASSOPTIONcaptionsoff
  \newpage
\fi

\footnotesize
\bibliographystyle{IEEEtran}
\bibliography{bibtex/bib/references}

\end{document}